\input{aipcheck}

\documentclass[
    ,final            
  ]
  {aipproc}

\layoutstyle{6x9}

\begin{document}

\title{Pion Electroproduction and VCS at the $\Delta$ Resonance Region }



\author{Nikolaos Sparveris }{
  address={Department of Physics, Temple University, Philadelphia, PA 19122, USA}
}

\begin{abstract}

 The study of the $\gamma^* N\rightarrow \Delta$ reaction presents the
 best quantitative method to explore the deviation of hadron shapes from
 spherical symmetry (nonspherical amplitudes). Significant non-spherical
 electric (E2) and Coulomb quadrupole (C2) amplitudes have been observed with
 good precision as a function of $Q^2$ from photon point up to
 $\approx 7 ~(GeV/c)^2$. Quark model calculations for these quadrupole
amplitudes are at least an order of magnitude too small and even
have the wrong sign. Lattice QCD, chiral effective field theory, and
dynamic model calculations which include the effects of the
pion-cloud are in approximate agreement with experiment.

\end{abstract}

\maketitle


\section{INTRODUCTION}

  The complex quark-gluon and meson cloud dynamics of hadrons
  give rise to non-spherical components in their wavefunction which in a
  classical limit and at large wavelengths will correspond to a "deformation".
  The spectroscopic quadrupole moment provides the most reliable and interpretable
  measurement of these components. For the proton, the only stable hadron, it
  vanishes identically because of its spin 1/2 nature. Instead, the signature
  of the non-spherical components of the proton is sought in the presence of
  resonant quadrupole amplitudes $(E^{3/2}_{1+}, S^{3/2}_{1+})$ in the
  predominantly magnetic dipole ($M^{3/2}_{1+}$)  $\gamma^* N\rightarrow \Delta$
  transition. Nonvanishing resonant quadrupole amplitudes will signify that
  either the proton or the $\Delta^{+}(1232)$ or more likely both are deformed.
  The ratios CMR $= Re(S^{3/2}_{1+}/M^{3/2}_{1+})$  and EMR $= Re(E^{3/2}_{1+}/M^{3/2}_{1+})$
  are routinely used to present the relative magnitude of the amplitudes of interest.

  In the quark model, the non-spherical amplitudes in the nucleon and $\Delta$ are caused by
  the non-central, tensor interaction between quarks \cite{glashow}. However, the magnitudes
  of this effect for the predicted E2 and C2 amplitudes \cite{capstick} are at least an
  order of magnitude too small to explain the experimental results and even the dominant M1 matrix element is
  $\simeq$ 30\% low. A likely cause of these dynamical shortcomings
  is that the quark model does not respect chiral symmetry, whose spontaneous breaking leads
  to strong emission of virtual pions (Nambu-Goldstone Bosons)\cite{amb}. These couple to
  nucleons as $\vec{\sigma}\cdot \vec{p}$ where  $\vec{\sigma}$ is the nucleon spin, and
  $\vec{p}$ is the pion momentum. The coupling is strong in the p wave and mixes in non-zero
  angular momentum components. Based on this, it is physically reasonable to expect that
  the pionic contributions increase the M1 and dominate the E2 and C2 transition matrix
  elements in the low $Q^2$ (large distance) domain. This was first indicated by adding
  pionic effects to quark models\cite{quarkpion1,quarkpion2,quarkpion3}, subsequently
  in pion cloud model calculations\cite{sato_lee,dmt}, and recently demonstrated in
  effective field theory (chiral) calculations \cite{pasc}.

\section{EXPERIMENTAL AND THEORETICAL LANDSCAPE}

 In recent years an extensive experimental and theoretical effort has been focused on
 identifying and understanding the origin of possible non-spherical components in the
 nucleon wavefunction \cite{review} through the study of the $\gamma^* N\rightarrow \Delta$
 transition. The exploration of the two pion excitation channels has allowed the measurement
 of the E2 and C2 amplitudes up to $Q^2 \approx7~(GeV/c)^2$ \cite{joo,frol,pos01,spaprl,elsner,ungaro,spaplb,aznauryan,villano}
 with JLab dominating the intermediate
 and high momentum transfer region and with MAMI and Bates focusing at the low momentum transfers,
 a region where the pionic contributions are expected to dominate. Both quadrupole amplitudes
 have been precisely measured to be non zero, with EMR exhibiting a relatively constant behavior
 as a function of the momentum transfer while CMR is following a fall off as a function of $Q^2$.

 With the existence of non-spherical components in the nucleon wavefunction well established, recent
  investigations have focused on understanding the various mechanisms that could
  generate it.
  The experimental results \cite{joo,frol,pos01,spaprl,elsner,ungaro,spaplb,aznauryan,villano} are in reasonable agreement
  with predictions of models invoking the presence of non-spherical amplitudes and in
  strong disagreement with all nucleon models that assume sphericity for the proton and
  the $\Delta$. A wide range of theoretical approaches has
  been developed to interpret the experimental data such as the phenomenological model MAID~2007 \cite{mai00,kama},
  the dynamical calculations of Sato-Lee \cite{sato_lee} and of DMT
\cite{dmt} and the ChEFT calculation of Pascalutsa and Vanderhaegen
\cite{pasc}. The MAID model  which offers a flexible phenomenology
provides an overall consistent agreement with the experimental
results. The DMT and Sato-Lee are dynamical reaction models which
include pion cloud effects and both calculate the resonant channels
from dynamical equations. DMT uses the background amplitudes of MAID
with some small modifications while Sato-Lee calculate all
amplitudes consistently within the same framework with only three
free parameters. DMT exhibits a reasonable agreement with the data
while the Sato-Lee model on the other hand exhibits a clear
disagreement with the $\pi^+$ data which is a bit surprising if we
consider the reasonable agreement of the Sato-Lee calculation with
the $\pi^\circ$ channel experimental results \cite{spaprl}. The
chiral effective field theory (ChEFT) calculation of Pascalutsa and
Vanderhaegen is a systematic expansion based on QCD\cite{pasc}. The
results of this expansion up to next to leading order exhibit an
overestimation of the $\pi^+$ channel results which indicates that
the next order calculation is required. It is worth pointing out
that this calculation is in reasonable agreement with the
experimental results for the $\pi^\circ$ channel
\cite{stave,spaplb}.

\section{FUTURE PROSPECTS}

Although the Electric quadrupole amplitude has been sufficiently
mapped as a function of momentum transfer the Coulomb quadrupole
still needs to be further explored at the low $Q^2$ regime.
Jefferson Lab Hall A experiment E08-010 \cite{halla} will extend the
Coulomb quadrupole measurements lower in momentum transfer down to
$Q^2=0.04~(GeV/c)^2$ and it will map the low $Q^2$ region with
measurements of unprecedented precision. Data taking of the E08-010
experiment was completed in 2011 and the results are expected in
early 2013.

Another valuable prospect will be offered by the exploration of the
VCS excitation channel at the $\Delta$ region. A MAMI experiment
\cite{mami} will offer the first precise exploration of the weak
H$(e,e^\prime p)\gamma$ channel at the $\Delta$ at
$Q^2=0.20~(GeV/c)^2$. The experiment will offer the first results
for the quadrupole amplitudes through the photon channel; the
different nature of this reaction channel, being purely
electromagnetic, and the fact that it will be measured
simultaneously with the dominant $\pi^\circ$ channel \cite{stave},
will allow important tests of the reaction framework and of the
systematic uncertainties of the extracted resonant amplitudes. The
experiment will also offer a simultaneous measurement of the protons
electric generalized polarizability exploring further its non
trivial behavior at low momentum transfers. The experiment will
acquire data at the end of 2012.

The upcoming results from JLab and MAMI along with the expected
progress on the theoretical front will provide a definitive
measurement of the non-spherical components in the nucleon wave
function, the identification of the dynamics that give rise to it,
and will have a profound impact on our understanding of hadrons and
QCD in the confinement regime.



\bibliographystyle{aipproc}   

\bibliography{sample}



\end{document}